\documentclass[a4paper]{JACoW-GSI}
\pdfoutput=1

\usepackage{graphicx}
\usepackage{url}

\usepackage{subfigure}
\usepackage[sort&compress, sort, numbers]{natbib}

\newcommand{\prt}[1]{\ensuremath{\mathrm{#1}}}
\newcommand{\Dz}{\prt{D^0}}
\newcommand{\Dzbar}{\prt{\bar{D}^0}}
\newcommand{\Ds}{\prt{D_s}}
\newcommand{\Dsp}{\prt{D_s^+}}

\newcommand{\Dspm}{\prt{D_s^{\pm}}}
\newcommand{\Dp}{\prt{D^+}}

\newcommand{\Dpm}{\prt{D^{\pm}}}

\newcommand{\KstarZ}{\prt{K^{*0}}}
\newcommand{\KstarZbar}{\prt{\bar{K}^{*0}}}
\newcommand{\thet}{\ensuremath{\theta_{\eta}}}

\newcommand{\BF}{\ensuremath{\mathcal{B}}}
\newcommand{\E}[1]{\ensuremath{\cdot 10^{#1}}}

\def \cb{{\cal B}}
\def \eea{\end{eqnarray}}
\def \eeq{\end{equation}}

\def \ol{\overline}
\def \s{\sqrt{2}}
\def \st{\sqrt{3}}

\def \p{\prime}
\def \im{{\it{i}}}
\newcommand{\Kp}{\ensuremath{K^+}}
\newcommand{\Km}{\ensuremath{K^-}}
\newcommand{\KS}{\ensuremath{K_{S}^0}\,}
\newcommand{\pip}{\ensuremath{\pi^+}}
\newcommand{\pim}{\ensuremath{\pi^-}}
\newcommand{\piz}{\ensuremath{{\pi^0}}}

\newcommand{\BFKSK}{\ensuremath{1.49 \pm 0.07 \pm 0.05}}
\newcommand{\BFKKpi}{\ensuremath{5.50 \pm 0.23 \pm 0.16}}
\newcommand{\BFKKpipiz}{\ensuremath{5.65 \pm 0.29 \pm 0.40}}
\newcommand{\BFKSKpipi}{\ensuremath{1.64 \pm 0.10 \pm 0.07}}
\newcommand{\BFpipipi}{\ensuremath{1.11 \pm 0.07 \pm 0.04}}
\newcommand{\BFpieta}{\ensuremath{1.58 \pm 0.11 \pm 0.18}}
\newcommand{\BFpietaprime}{\ensuremath{3.77 \pm 0.25 \pm 0.30}}
\newcommand{\BFKpipi}{\ensuremath{0.69 \pm 0.05 \pm 0.03}}

\newcommand{\BRKSK}{\ensuremath{0.270 \pm 0.009 \pm 0.008}}
\newcommand{\BRKKpipiz}{\ensuremath{1.03 \pm 0.05 \pm 0.08}}
\newcommand{\BRKSKpipi}{\ensuremath{0.298 \pm 0.014 \pm 0.011}}
\newcommand{\BRpipipi}{\ensuremath{0.202 \pm 0.011 \pm 0.009}}
\newcommand{\BRpieta}{\ensuremath{0.288 \pm 0.018 \pm 0.033}}
\newcommand{\BRpietaprime}{\ensuremath{0.69\pm 0.04 \pm 0.06}}
\newcommand{\BRKpipi}{\ensuremath{0.125 \pm 0.009 \pm 0.005}}

\newcommand{\AKSK}{\ensuremath{+4.9 \pm 2.1 \pm 0.9}}
\newcommand{\AKKpi}{\ensuremath{+0.3 \pm 1.1 \pm 0.8}}
\newcommand{\AKKpipiz}{\ensuremath{-5.9 \pm 4.2 \pm 1.2}}
\newcommand{\AKSKpipi}{\ensuremath{-0.7 \pm 3.6 \pm 1.1}}
\newcommand{\Apipipi}{\ensuremath{+2.0 \pm 4.6 \pm 0.7}}
\newcommand{\Apieta}{\ensuremath{-8.2 \pm 5.2 \pm 0.8}}
\newcommand{\Apietaprime}{\ensuremath{-5.5 \pm 3.7 \pm 1.2}}
\newcommand{\AKpipi}{\ensuremath{+11.2 \pm 7.0 \pm 0.9}}

\newcommand{\un}[2]{\ensuremath{#1\,\mathrm{#2}}}

\newcommand{\cleo}{\mbox{CLEO-c}}
\newcommand{\Cleo}{\mbox{CLEO-c}}

\newcommand{\Figref}[1]{Figure~\ref{#1}}
\newcommand{\figref}[1]{Fig.~\ref{#1}}

\newcommand{\tabref}[1]{Tab.~\ref{#1}}

\begin{document}
\title{Non-leptonic \Dz, \Dp, and \Dsp\ Branching Fractions}

\author[1]{Jonas Rademacker on behalf of the \cleo\ collaboration}
\affil[1]{University of Bristol, UK}

\maketitle

\section{Introduction}

Non-leptonic charm decays provide insights into both electro-weak and
strong dynamics. This includes the study of long-distance hadronic
effects, the approximate symmetries of strong interactions, and
precision tests of the Standard Model.

In these proceedings we summarise recent results in non-leptonic
branching fraction measurements of \Dz, \Dpm, and \Dspm\ mesons,
including measurements of relative and absolute branching fractions in
inclusive and exclusive modes, radiative decays, and measurements of
direct CP violation. Other aspects of hadronic charm decays are
covered elsewhere in these proceedings~\cite{Wilkinson:2009wx,
  DosReis:2009rv, MixBaBar, MixBELLE, MixCLEO, MixCDF, MixHFAG,
  MixTheory1, Bigi:Charm09}.

\section{Charm decays to two pseudoscalars}
\begin{figure}[b]
\includegraphics[width=0.95\columnwidth]{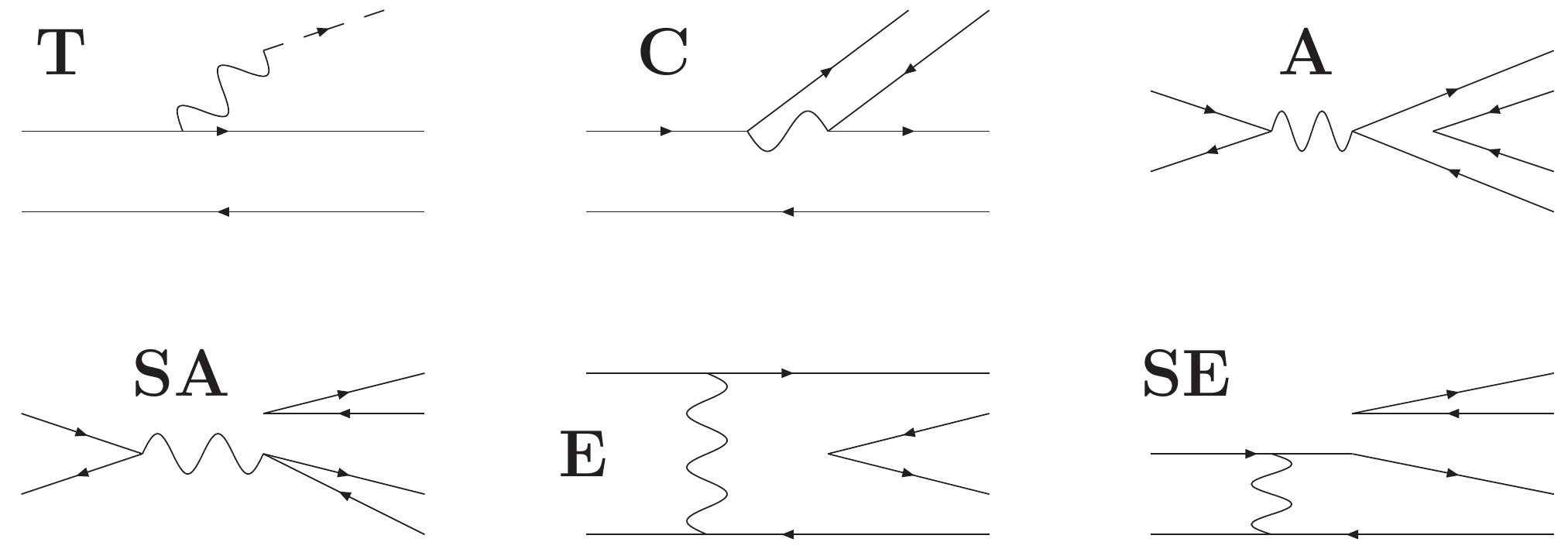}
\caption{Quark flow diagrams used in the analysis of \cleo's \prt{D\to PP}
  data~\cite{cleoDPP2009} by
  Bhattacharya~\&~Rosner~\cite{Bhattacharya:2009ps}:
  \textbf{T}ree, \textbf{C}olour-suppressed tree,
  \textbf{A}nnihiliation, \textbf{S}inglet-emission with \textbf{A}nnihilation,
  \textbf{E}xchange, and \textbf{S}inglet-emission with \textbf{E}xchange.
\label{fig:quark_diagrams}}
\end{figure}
\begin{figure}[bh]
\parbox{0.3\textwidth}{
\includegraphics[width=0.3\textwidth]{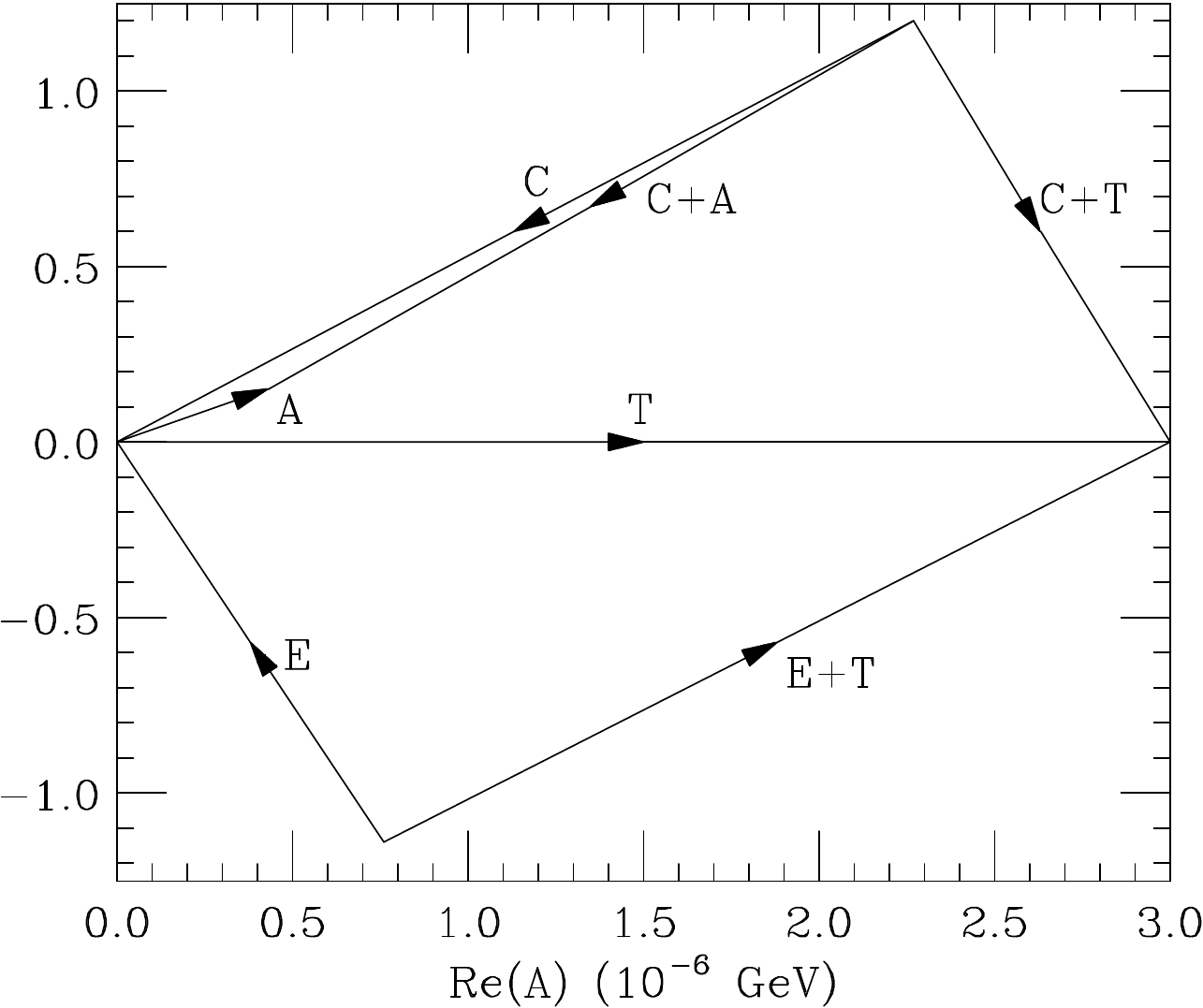}}
\caption{Construction of topological amplitudes
  in the complex plane based on \cleo's recent
  measurements~\cite{cleoDPP2009} of for the solution
  with $\thet=11.7^o$, reproduced from~\cite{Bhattacharya:2009ps}.
  \label{fig:RosnersArrows_12deg}}
\end{figure}
\cleo\ has recently published the results of branching fractions of
\Dz, \Dp, and \Ds\ decays to two pseudoscalars, based on an analysis
of \cleo's full data set~\cite{cleoDPP2009}, with \un{818}{pb^{-1}} at
$\psi(3770)$ corresponding to 3\E{6} \prt{D^0\bar{D}^0} pairs and
2.4\E{6} \prt{D^+D^-} pairs; and \un{586}{pb^{-1}} at $\sqrt{s} =
\un{4170}{MeV}$ corresponding to 5.3\E{5} \prt{D_s^{\pm} D_s^{\mp *}}
pairs. Many of the resulting branching fraction measurements are more
precise than the previous world average~\cite{pdg2008}, and some decay
modes have been seen for the first time. These results are summarised
in \tabref{tab:DPP_results} on page~\pageref{tab:DPP_results}. In the
table, as in the rest of this paper, the mention of one decay process
always implies also the charge-conjugate process, and if a number is
given with two uncertainties, the first refers to the statistical and
the second to the systematic uncertainty.
Bhattacharya~\&~Rosner~\cite{Bhattacharya:2009ps} have analysed these
results in terms of the diagrammatic approach~\cite{Diagrams_0,
  Diagrams_1, Diagrams_2, Diagrams_3}. The decay amplitudes are
expressed in terms of topological quark-flow diagrams; the diagrams
used in this analysis are given in \figref{fig:quark_diagrams}. 
\begin{figure*}
\begin{center}
\caption{Branching ratios and quark-flow diagram amplitudes for Cabibbo-favored
decays of charmed mesons to a pair of pseudoscalars with 2 different values of
$\thet$, reproduced from~\cite{Bhattacharya:2009ps}.
\label{tab:CF2}}
\begin{tabular}{c l c c c c}
\hline \hline
Meson & Decay & $\cb$ \cite{cleoDPP2009} & Rep.\ & \multicolumn{2}{c}{Predicted $\cb$ ($\%$)} \\
      & mode  & ($\%$) &  & $\thet = 11.7^\circ$ & $\thet= 19.5^\circ$    \\ \hline
$D^0$  &$K^- \pi^+$       &3.891$\pm$0.077&  $T+E$       &3.891&3.905\\
       &$\ol{K}^0 \pi^0$  &2.380$\pm$0.092& $(C-E)/\s$   &2.380&2.347\\
       &$\ol{K}^0 \eta$   &0.962$\pm$0.060&~$\frac{C}{\s}\sin(\thet+\phi_1)
       -\frac{\st E}{\s}\cos(\thet+2\phi_1)$&0.962&1.002\\
       &$\ol{K}^0 \eta^\p$&1.900$\pm$0.108&-$\frac{C}{\s}\cos(\thet+\phi_1)
       -\frac{\st E}{\s}\sin(\thet+2\phi_1)$&1.900&1.920\\\hline
$D^+$  &$\ol{K}^0 \pi^+$  &3.074$\pm$0.097& $C+T$        &3.074&3.090\\ \hline
$D^+_s$&$\ol{K}^0 K^+$    &2.98$\pm$0.17  & $C+A$        &2.980&2.939\\
       &$\pi^+ \eta$      &1.84$\pm$0.15  & $T\cos(\thet+\phi_1)
       -\s A\sin(\thet+\phi_1)$ &1.840&1.810\\
       &$\pi^+ \eta^\p$   &3.95$\pm$0.34  & $T\sin(\thet+\phi_1)
       +\s A\cos(\thet+\phi_1)$ &3.950&3.603\\ \hline \hline
\end{tabular}
\end{center}
\end{figure*}
These are not to be confused with Feynman diagrams. The amplitude
represented by each diagram includes the contributions from the weak
and the strong interaction, to all orders, including long-distance
effects.  Flavour symmetries of the strong interaction are used to
express different \Dz, \Dpm\ and \Dspm\ two-body decay amplitudes in
terms of the same set of six diagrams.  Note that, because the
amplitudes associated to each diagram include final state interaction,
the amplitudes established from two body decays do not predict
amplitudes for decays to three or more particles in the final
state. The expressions for Cabibbo-favoured (CF) decays in terms of
these diagrams are given in \tabref{tab:CF2}. The singlet
contributions to these decays are deemed to be negligible. The table
compares the measured branching fractions with the result from the
best-fit to the quark flow diagram formalism for two solutions.  One
where the octet-singlet mixing angle $\thet$ is fixed to
$\thet=\arcsin(1/3)=19.5^o$, and another, where $\thet$ is allowed to
vary, giving $\thet=11.7^o$. The latter case has as many parameters as
there are CF decay rates used as constraints, so the agreement between
the prediction from the formalism given in the fifth column of
\tabref{tab:CF2}, and the measured CF amplitudes given in the second,
is exact by construction. A further solution, with $|T| < |C|$, is
also discussed in the paper~\cite{Bhattacharya:2009ps}.
\Figref{fig:RosnersArrows_12deg} shows the construction of the
amplitudes from the rates given in \tabref{tab:CF2} for the case for
the $\thet=11.7^o$ case. The numerical values are
\begin{eqnarray*}
T &=&  3.003\pm0.023 \\
C &=& (2.565\pm0.030)\,\exp\left[\im(-152.11\pm0.57)^\circ\right] \\
E &=& (1.372\pm0.036)\,\exp\left[\im( 123.62\pm1.25)^\circ\right] \\
A &=& (0.452\pm0.058)\,\exp\left[\im(  19^{+15}_{-14})^\circ\right]
\end{eqnarray*}
These results are then used to predict the decay amplitudes of singly
Cabibbo suppressed (SCS) and doubly Cabibbo suppressed (DCS) two body
decays by assuming that the SCS (DCS) amplitudes are the CF
amplitudes, scaled by a factor $\lambda=\sin\theta_c$ ($\lambda^2 =
\sin^2\theta_c$) where $\theta_c$ is the Cabibbo angle. The
predictions for decays involving kaons and pions only are mostly in
reasonable agreement with measurement although the approach
considerably overestimates \prt{\BF\left(D^0 \to \pi^+ \pi^-\right)}
and underestimates \prt{\BF\left(D^0\to K^+ K^-\right)}. For SCS decays
involving $\eta$ and $\eta^{\prime}$, there are indications for a
non-negligible contribution from the singlet annihilation diagram.

A detailed description of this approach and its result can be found
in~\cite{Bhattacharya:2009ps} and \cite{Diagrams_0, Diagrams_1,
  Diagrams_2, Diagrams_3}. A comprehensive review of hadronic charm
decays and their analysis using this and other methods can be
found in~\cite{RydPetrov_bigOverview:2009uf}.

\section{\prt{K^0}, \prt{\bar{K}^0} interference}
\begin{figure}
\includegraphics[width=0.95\columnwidth]{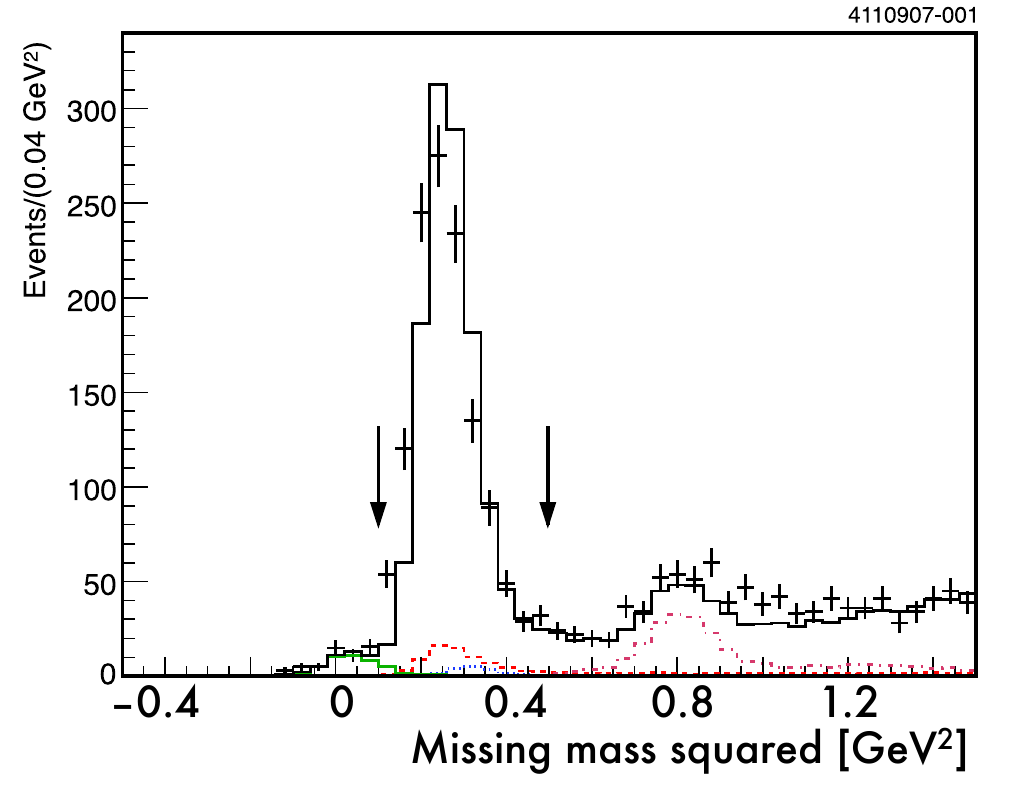}
\caption{The missing mass distribution in the reconstruction of
  \prt{\Dz \to K_L \pi^0} at \cleo~\cite{CLEO:KKbarInterference}.  The points
  with error bars are data, and the solid line is a Monte Carlo
  simulation. The dashed, colored lines represent simulations of the
  peaking backgrounds. The difference in the peak position is
  understood and due to a minor discrepancy in the calorimeter
  simulation at large photon energies.\label{fig:CLEO:KKbarInterference}}
\end{figure}

As pointed out by Bigi~\&~Yamamoto~\cite{Bigi1995363}, the decay rates
of \prt{\Dz \to K_S\pi^0} and \prt{\Dz \to K_L\pi^0} are not the same
because of the interference of the CF component \prt{\Dz \to \bar{K}^0
  \pi^0} with the DCS \prt{\Dz \to {K}^0 \pi^0} component which enters
with a different sign for decays to \prt{K_L} and \prt{K_S}:
{\small
\begin{eqnarray*}
 A\left(\prt{\Dz \to K_S\pi^0}\right) & = & 
 A\left(\prt{\Dz \to \bar{K}^0 \pi^0}\right) +  
 A\left(\prt{\Dz \to K^0 \pi^0}\right)
\\
 A\left(\prt{\Dz \to K_L\pi^0}\right) & = & 
 A\left(\prt{\Dz \to \bar{K}^0 \pi^0}\right) -
 A\left(\prt{\Dz \to K^0 \pi^0}\right) 
\end{eqnarray*}
}
The amplitudes $A\left(\prt{\Dz \to \bar{K}^0 \pi^0}\right)$ and
$A\left(\prt{\Dz \to K^0 \pi^0}\right)$ are related by an interchange
of $u$ and $s$ quarks. Assuming U-spin symmetry of the strong
interaction, the decay rate asymmetry is given given
by~\cite{Bigi1995363}:
\begin{eqnarray*}
A_{K_{S,L}\pi^0} &=& 
\frac{\Gamma\left(\Dz \to K_S \pi^0\right) - 
  \Gamma\left(\Dz \to K_L \pi^0\right) }{
  \Gamma\left(\Dz \to K_S \pi^0\right) +
  \Gamma\left(\Dz \to K_L \pi^0\right) }
\\ &=& 2\tan^2{\theta_c} = 0.109
\end{eqnarray*}
A measurement of $A_{K_{S,L}\pi^0}$ therefore provides a test of
U-spin symmetry, which is important for example for extracting the
CP-violating parameter $\gamma$ from \prt{B_s \to KK} and \prt{B_d \to
  \pi\pi} decays~\cite{Fleischer:1999pa, Fleischer:2000un}. The
reconstruction of \prt{D^0 \to K_L \pi^0} is challenging because it
involves two neutral particles. \cleo\ uses its CsI calorimeter to
identify the \prt{\pi^0}. The four-momentum of the practially
invisible \prt{K_L} is reconstructed using beam constraints,
benefiting from the very clean environment at \cleo\ where the
\prt{D\bar{D}} pairs produced absorb the entire beam energy, with no
underlying event. The resulting missing mass-squared distribution is
shown in \figref{fig:CLEO:KKbarInterference}.
The assymmetry, measured in \un{281}{pb^{-1}} of data,
is~\cite{CLEO:KKbarInterference}:
\[
A_{K_{S,L}\pi^0} = 0.108 \pm 0.025 \pm 0.024,
\]
which is in excellent agreement with the prediction
by~\cite{Bigi1995363} based in U-spin symmetry.
Theoretical prediction for the related asymmetry
\begin{eqnarray*}
A_{K_{S,L}\pi^+} &=& 
\frac{\Gamma\left(\Dp \to K_S \pi^+\right) - 
  \Gamma\left(\Dp \to K_L \pi^+\right) }{
  \Gamma\left(\Dp \to K_S \pi^+\right) +
  \Gamma\left(\Dp \to K_L \pi^+\right) }
\end{eqnarray*}
 are more difficult as there is no such clean symmetry. Using SU(3),
 Gao predicts~\cite{Gao:2006nb}
\(A_{K_{S,L}\pi^+} \approx 0.04\). Based on the diagrammatic approach,
Bhattacharya~\&~Rosner~\cite{Bhattacharya:2009ps} predict 
\(A_{K_{S,L}\pi^+} = - 0.005 \pm 0.013\). Both are consistent with
\cleo's measurement~\cite{CLEO:KKbarInterference} of 
\[
A_{K_{S,L}\pi^+} = 0.022 \pm 0.016 \pm 0.018.
\]

\section{Decays to vector-mesons and $\eta$}

BaBar analysed \un{467}{fb^{-1}} of data corresponding to about 1
billion D mesons. Preliminary results for the decay \prt{\Dz \to
  V\eta}, where \prt{V=\omega, \phi, \KstarZ}, have been presented at
the APS April meeting 2009~\cite{maloneAPS}. The \prt{\omega\eta} and
\prt{\KstarZ\eta} mode have been observed for the first time.  These
results, and a previous measurement by
BELLE~\cite{BELLEphiGamma}, 
are summarised in \tabref{tab:Veta}. The measurements are compared to
predictions by Bhattacharya~\&~Rosner~\cite{RosnerVeta}, who use the
same diagrammatic approach that has been discussed earlier. This
yields two solutions, of which solutions A is preferred. While
Bhattacharya~\&~Rosner~\cite{RosnerVeta} show in their paper that the
predictions based on the diagrammatic approach agree well for many
\Dz\ to vector-pseudoscalar decays, there are significant differences
for two of the decays shown in \tabref{tab:Veta}.
\begin{table*}
\begin{center}
\caption{Recent results for \prt{\Dz\to V\eta}. The BaBar results
  shown in this table are preliminary.\label{tab:Veta}}
\begin{tabular}{|| l | c | c | r  l||}
\hline
{\small Channel}
& \multicolumn{2}{c|}{Prediction/$10^{-3}$~\cite{RosnerVeta}}
& \multicolumn{2}{c||}{Measured $\BF/10^{-3}$/} \\
\prt{\Dz \to } & Sol A & Sol B &  & \\
\hline
\prt{\phi\eta } 
& $0.93 \pm 0.09$ & $1.4\pm 0.1$ 
& $0.14\pm 0.04$ & (BELLE~\cite{BELLEphiGamma}) \\
&&&
$0.21 \pm 0.01 \pm 0.02$ & (BaBar-prelim~\cite{maloneAPS})
\\\hline
\prt{\omega\eta}
& $1.4 \pm 0.09$ & $1.27\pm 0.09$ 
& $2.21 \pm 0.08 \pm 0.22$ & (BaBar-prelim~\cite{maloneAPS})
\\\hline
\prt{\KstarZ\eta}
& $0.038 \pm 0.004$ & $0.037\pm 0.004$ 
& $0.048\pm 0.010 \pm 0.004$ & (BaBar-prelim~\cite{maloneAPS})
\\\hline
\end{tabular}
\end{center}
\end{table*}

\section{Radiative decays}

In 2008, BaBar reported the first observation of the decay \prt{\Dz
  \to \KstarZbar \gamma}~\cite{BaBarVgamma2008}, and an improved
measurement of \prt{\Dz \to \KstarZbar \gamma}. 
\cleo\ has since been
able to confirm the observation of \prt{\Dz \to \KstarZbar \gamma}.
\begin{table}
\caption{Recent results for radiative \Dz\ decays. The \cleo\ results
  shown in this table are preliminary.}
{
\begin{tabular}{|| l | l  p{2.5cm}||}
\hline
\mbox{\rotatebox{70}{\small Channel}}& \multicolumn{2}{c||}{$\BF$ranching Fraction} \\
\hline
\prt{\overline{K}^*\! \gamma } 
& \(\left(3.22 \pm 0.20 \pm 0.27\right)\cdot 10^{-4}\) &
(BaBar~\cite{BaBarVgamma2008}) \\
& $\left( 4.37 \pm 0.37 \pm 0.52 \right) \cdot 10^{-4}$ & \small (\cleo\ prel~\cite{Klein:2008zz})
\\\hline
\prt{\phi \gamma } & 
$\left(2.6^{+0.70}_{-0.61}\;\;\mbox{}^{+0.15}_{-0.17}\right)\cdot 10^{-5}$
& (BELLE~\cite{BELLEphiGamma})\\
&
\(\left(2.73 \pm 0.30 \pm 0.26\right) \cdot 10^{-5}\)
& (BaBar~\cite{BaBarVgamma2008}) \\
&
$\left( 2.21 \pm 0.95 \pm 0.28 \right) \cdot 10^{-5}$ & \small (\cleo\ prel~\cite{Klein:2008zz})
\\\hline
\prt{\gamma \gamma } & $ < 8.93\cdot 10^{-6} (90\% CL)$ &
\small (\cleo\ prel~\cite{Klein:2008zz})\\\hline
\prt{\rho \gamma } & $ < 3.63\cdot 10^{-5} (90\% CL)$ & \small
(\cleo\ prel~\cite{Klein:2008zz})\\\hline
\prt{\omega \gamma } & $ < 3.00\cdot 10^{-5} (90\% CL)$ &
\small (\cleo\ prel~\cite{Klein:2008zz})\\\hline
\end{tabular}\\
}
\end{table}
In contrast to radiative B decays, radiative charm decays are
dominated by long-distance contribution.
One way to describe radiative decays is the Vector Meson Dominance
(VMD) approach. The radiative decay is assumed to proceed
predominantly via an off-shell $\rho^0$ that then annihilates into a
photon, giving the following relationship between the decay
amplitudes~\cite{VMDTheory}
\[
 A(\prt{\Dz\to V\gamma}) = (e/f_{\rho}) A(\Dz \to V \rho^0)
\]
where $A(\Dz \to V \rho^0)$ needs to be calculated taking into account
that the $\rho$ is off-shell. This predicts for the ratio of decays rates:
\[
\frac{\BF(\Dz \to \phi\gamma)}{\BF(\Dz \to \KstarZbar \gamma)}
\approx \frac{\BF(\Dz \to \phi\rho)}{\BF(\Dz \to \KstarZbar \rho)}
\]
This is in fact the case:
\begin{eqnarray*}
 \frac{\BF(\Dz \to \phi\gamma)}{\BF(\Dz \to \KstarZbar \gamma)}
 &=& \left(6.27 \pm 0.71 \pm 0.79\right) \cdot 10^{-2}
\;\;\mbox{\cite{BaBarVgamma2008}}
\\
 \frac{\BF(\Dz \to \phi\rho)}{\BF(\Dz \to \KstarZbar \rho)}
 &=&  \left(6.7 \pm 1.6\right) \cdot 10^{-2}
\;\;\mbox{\cite{pdg2008}}
\end{eqnarray*}
However, using $(e/f_{\rho}) = 0.06$~\cite{VMD2}, one would also
expect
\[
\frac{\BF(\prt{\Dz\to V\gamma})}{\prt{\Dz \to V \rho^0}} \approx 0.0036
\]
but the measured ratio for $V=\KstarZbar$ as well as $V=\phi$ is
\[
\frac{\BF(\Dz \to V \gamma)}{\BF(\Dz \to V\rho^0)} \approx 0.02 
\;\;\;\mbox{for V=\prt{\KstarZbar} or \prt{\phi}}
\]
which is about a factor of $6$ larger than expected from the VMD
approach.

\section{\prt{\Dz \to p\bar{n}}}
\begin{figure}
\includegraphics[width=0.95\columnwidth]{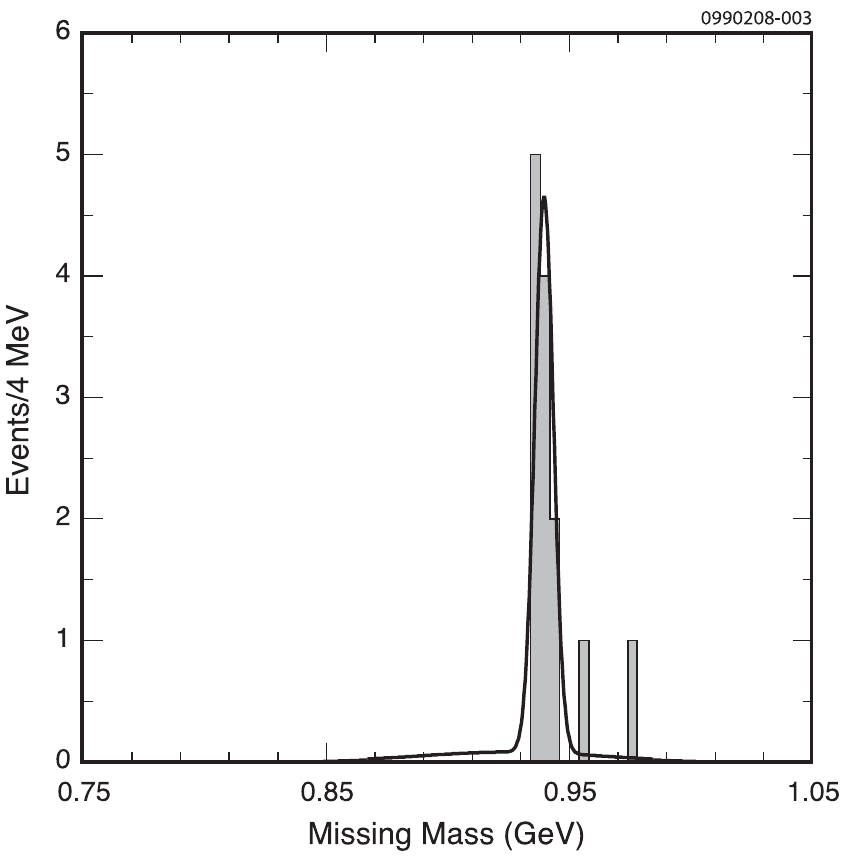}
\caption{The missing mass of the $n$ in \prt{\Dsp \to p\bar{n}} decays,
  at \cleo~\cite{CLEOpn}.\label{fig:pn}}
\end{figure}
\begin{figure}[h]
\includegraphics[width=0.95\columnwidth]{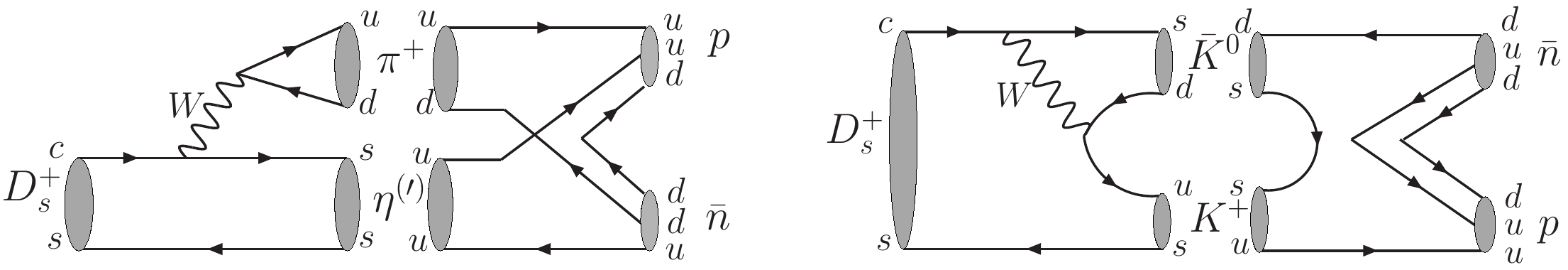}
\caption{Long distance effects dominate the decay \prt{\Dsp \to
    p\bar{n}}.\label{fig:pnLD}}
\end{figure}

The first observation of a meson decaying to two baryons has been made
by \cleo\ in the mode \prt{\Dsp \to p \bar{n}}, which is also the only
kinematically allowed baryonic decay of a light charm meson (\Dz, \Dp,
or \Ds). \Cleo\ reconstruct the anti-neutron from the missing mass
with virtually no background, as shown in \figref{fig:pn}. \cleo\
measures the follwing branching fraction~\cite{CLEOpn}:
\[
\BF(\Dsp \to p\bar{n}) 
= \left(1.30 \pm  0.36^{+0.12}_{-0.16}\right) \cdot 10^{-3}
\]
This decay mode is dominated by long-distance effects as those shown
in \figref{fig:pnLD}.
Chen, Cheng and Hsio~\cite{pnTheory} estimate these as
\(\BF(\Dsp \to p\bar{n}) \approx
\left(0.8^{+2.4}_{-0.6}\right) \cdot 10^{-3}\) in agreement with
\cleo's observation - short-distance contributions from the
annihilation diagram are about 3 orders of magnitude smaller.

\section{Absolute Branching Fractions}
\begin{figure}
\includegraphics[width=0.97\columnwidth]{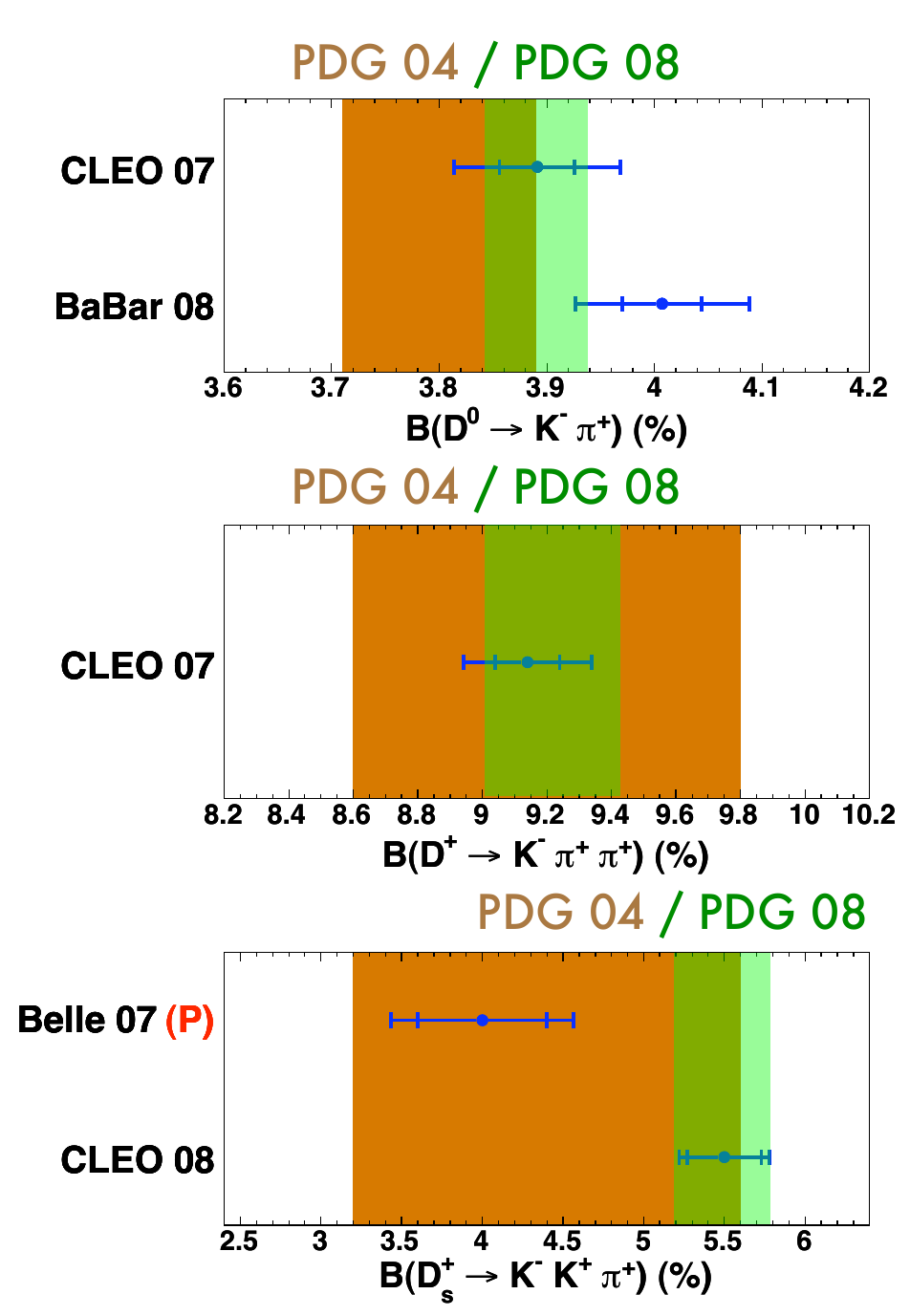}
\caption{Impact of recent absolute Branching ration measurements on
  the PDG average. The brown bar represents the PDG average in 2004,
  while the semi-transparent green bar (that partially overlaps with
  the brown bar) represents the status in 2008. Recent measurements by
  BaBar~\cite{absBaBar},  BELLE~(preliminary~\cite{absBELLE}) and
  \cleo~\cite{absCLEODz, absCLEODs} are represented by blue lines
  with error bars.\label{fig:absBars}}
\end{figure}
\begin{figure}
\includegraphics[width=0.95\columnwidth]{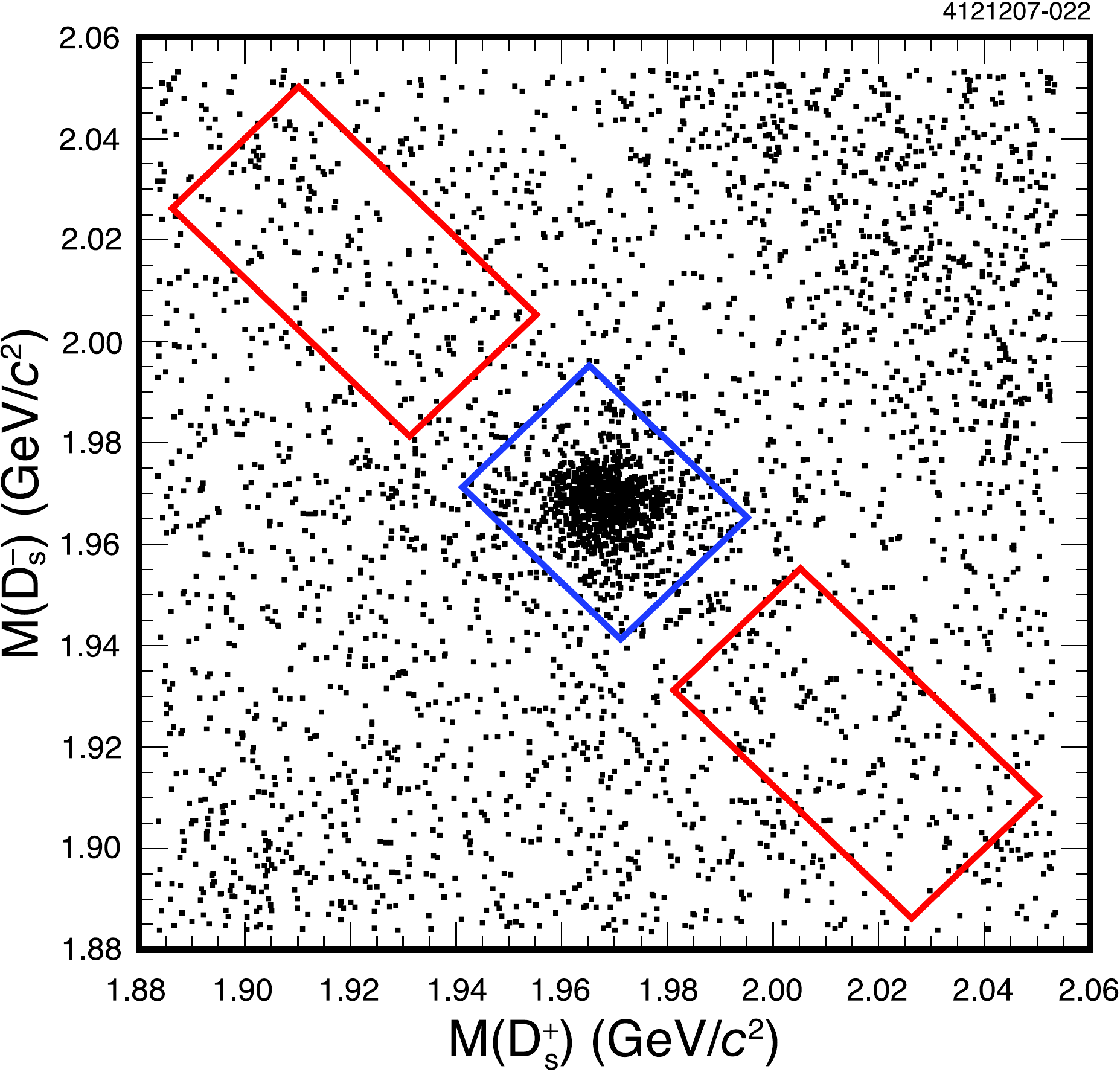}
\caption{Invariant mass of \Ds\ pairs reconstructed at
  \cleo~\cite{absCLEODs}.
\label{fig:DsScatter}}
\end{figure}
Absolute Branching fraction measurements are particularly important
for those decays frequently used as normalisation modes.
 BaBar, BELLE
and \cleo\ published measurements of absolute branching fractions,
using different techniques:
\begin{itemize}
\item BaBar obtains a normalisation by reconstructing \prt{D^*\to
    D\pi} using only the slow pion in this decay chain, and
  information from the rest of the event, but not the \prt{D}
  itself~\cite{absBaBar}. 
\item \cleo\ produces charm mesons always in pairs, either \prt{e^+
    e^- \to \psi(3770) \to D\bar{D}} for \Dz\ or \Dpm, or \prt{e^+ e^-
    \to D^{\pm} D^{*\mp}}. One charm meson provides the normalisation
  for the decay rates of the other~\cite{absCLEODz,
    absCLEODs}. \Figref{fig:DsScatter} shows the invariant mass
  distribution of \Ds\ pairs at \cleo. In the \Dz-\Dzbar\ system,
  there are complications and very interesting physics arising from
  quantum correlations which are discussed elsewhere in these
  proceedings~\cite{Wilkinson:2009wx}.
\item BELLE uses the process \prt{e^+ e^- \to D^{*+}_s D^{-}_{s1}(\to
    \bar{D}^{*0}K^-)}, again, one charm meson provides the
  normalisation for the other~\cite{absBELLE}.
\end{itemize}
\Figref{fig:absBars} illustrates the significant increase in precision
achieved in recent years for the most important normalisation modes. %
The full set of \cleo's absolute \Ds\ branching fractions measurements
are given in \tabref{tab:absDs}, reproduced from~\cite{absCLEODs}.
\begin{table*}
  \caption{\label{tab:absDs}Results from \cleo's
    recent measurement of absolute \Ds\ branching
    fractions~\cite{absCLEODs}, the world average branching
    fractions before \cleo's measurement~\cite{pdg2006}, 
    ratios of branching fractions to
    $\BF(\Dsp\to\Km\Kp\pip)$, and charge asymmetries
    $\mathcal{A}_{CP}$.  Uncertainties on \cleo\ measurements are
    statistical and systematic, respectively. Table reproduced
    from~\cite{absCLEODs}}
\begin{center}
\begin{tabular}{lcccc}
\hline\hline
Mode & \cleo\ result \BF\ (\%)~\cite{absCLEODs} & PDG 2007 fit \BF\ (\%)~\cite{pdg2006} & $\BF/\BF(\Km\Kp\pip)$ & $\mathcal{A}_{CP}$ (\%)\\
\hline
\KS\Kp & \BFKSK & $2.2 \pm 0.4$ & \BRKSK & \AKSK \\
\Km\Kp\pip & \BFKKpi & $5.3 \pm 0.8$ & 1 & \AKKpi\\
\Km\Kp\pip\piz & \BFKKpipiz & --- & \BRKKpipiz & \AKKpipiz\\
\KS\Km\pip\pip & \BFKSKpipi & $2.7 \pm 0.7$ & \BRKSKpipi & \AKSKpipi\\
\pip\pip\pim & \BFpipipi & $1.24 \pm 0.20$ & \BRpipipi & \Apipipi\\
$\pip\eta$ & \BFpieta & $2.16 \pm 0.30$ & \BRpieta & \Apieta\\
$\pip\eta'$ & \BFpietaprime & $4.8 \pm 0.6$ & \BRpietaprime & \Apietaprime\\
\Kp\pip\pim & \BFKpipi & $0.67 \pm 0.13$ & \BRKpipi & \AKpipi\\
\hline\hline
\end{tabular}
\end{center}
\end{table*}

A frequently-used normalisation mode for \Ds\ branching ratios is the
decay \prt{\Ds \to \phi \pi}. This, however, is problematic because of
interference effects in the \prt{K^+K^-\pi^+} Dalitz plot, in
particular from \prt{f(980)}~\cite{Stone:2006bb,
  FocusDalitz,E687Dalitz}. \Cleo\ therefore publishes the absolute
branching fraction for \prt{\Dsp \to K^+ K^- \pi^+}, including the
entire phase space. However, when using this as a normalisation mode,
it can be advantageous to select events with a \prt{K^+ K^-} invariant
mass near the \prt{\phi} mass, in order to reject background. To
accommodate this, \cleo\ also publishes branching fractions for parts
of the \prt{\Dsp \to K^+ K^- \pi^+} phase space corresponding to
different cuts around the \prt{\phi} mass, but without making any
statement about the contribution of \prt{\Dsp \to \phi \pi} this
includes. The absolute \Ds\ branching fractions for different decay
modes from this analysis~\cite{absCLEODs} are given in \tabref{tab:absDs}.

\section{Inclusive \Ds\ branching fractions and exclusive \prt{D_s \to
    \omega X}}

In 2009, \cleo\ published a measurement the inclusive branching
fractions of \Ds\ in modes~\cite{CLEO:incDs:2009}, such as \prt{\Dsp
  \to \pi^+ X}, \prt{\Dsp \to \pi X}, etc, where \prt{X} stands for
any combination of particles. The results are reproduced in
\tabref{tab:incDs} on page \pageref{tab:incDs}.
While most inclusive branching fractions measured are compatible with
the sum of known exclusive rates~\cite{Gronau:2009vt}, this was
initially not the case for the inclusive branching fraction
\prt{\BF(\Ds \to \omega X)}, where $X$ stands for any combination of
particles. \cleo\ measures this to be $(6.1 \pm 1.4)\%$, far more than
the only known exclusive \prt{\omega} mode at the time,
$\prt{\BF\left(\Ds \to \pi^+ \omega\right)} = \left(0.25\pm
  0.09\right)\%$~\cite{pdg2008}.

Since then, \cleo\ has searched for the missing exclusive decay modes
to \prt{\omega}, and found them~\cite{CLEO:exclOmega:2009}. The
missing exclusive modes are mainly those were $X$ is two or three
pions. The full results are given in \tabref{tab:exclOmega}.
\begin{table}[h]
  \centering
  \caption{\label{tab:exclOmega}Branching fractions and upper
    limits for exclusive \Ds\ decays involving \prt{\omega},
    reproduced from~\cite{CLEO:exclOmega:2009}.}
  \begin{tabular}{l  c }
    \hline \hline
     Mode  &  $\mathcal{B}_\mathrm{mode}(\%)$ \\ \hline\hline
    \prt{D^+_s \rightarrow \pi^+ \omega}~~~~~~~~~~~~~~~~~~~~~~~~
    &$0.21 \pm 0.09 \pm 0.01$\\\hline
    \prt{D^+_s \rightarrow \pi^+ \pi^0 \omega}
    &$2.78 \pm 0.65 \pm 0.25$\\\hline
    \prt{D^+_s \rightarrow \pi^+ \pi^+ \pi^- \omega}
    &$1.58 \pm 0.45 \pm 0.09$\\\hline
    \prt{D^+_s \rightarrow \pi^+ \eta \omega}
    &$0.85 \pm 0.54 \pm 0.06$\\ 
    &$<2.13~(90\%~\mathrm{CL})$\\\hline
    \prt{D^+_s \rightarrow K^+ \omega}
    &$<0.24~(90\%~\mathrm{CL})$\\\hline
    \prt{D^+_s \rightarrow K^+ \pi^0 \omega}
    &$<0.82~(90\%~\mathrm{CL})$\\\hline
    \prt{D^+_s \rightarrow K^+ \pi^+ \pi^- \omega}
    &$<0.54~(90\%~\mathrm{CL})$\\\hline
    \prt{D^+_s \rightarrow K^+ \eta\omega}
    &$<0.79~(90\%~\mathrm{CL})$\\
   \hline\hline
  \end{tabular}
\end{table}

\section{Direct CP violation}

CP violation in charm decays provides one of the most sensitive probes
for New Physics, and we are only now reaching the sensitivity to
exploit this opportunity. In this article, we restrict ourselves to
the discussion of direct CP violation; CP violation in the
interference between mixing and decay is discussed elsewhere in these
proceedings~\cite{MixBaBar, MixBELLE, MixCLEO, MixCDF, MixHFAG,
  MixTheory1, Bigi:Charm09}.

CP violation in the charm decays in the SM is expected to be small, at
a level $<10^{-3}$~\cite{small_SMCPV_1, small_SMCPV_2,
  small_SMCPV_3}. However, new Physics can significantly enhance CP
asymmetries, especially in the case of singly Cabibbo suppressed
decays, which are sensitive to new contributions from QCD penguin
operators. This could yield direct CP violating effects of
$\mathcal{O}(10^{-2})$~\cite{CPV_in_SC_theory, Bianco:2003vb}.

The most precise measurements of direct CP asymmetries
\[
A_{CP} \equiv 
\frac{\Gamma\left(\prt{D \to f}\right) - 
      \Gamma\left(\prt{\bar{D} \to \bar{f}}\right)}{
      \Gamma\left(\prt{D \to f}\right) +
      \Gamma\left(\prt{\bar{D} \to \bar{f}}\right)}
\]
in SCS decays exist for the modes \prt{\Dz \to K^+ K^-} 
and \prt{ \Dz \to \pi^+ \pi^-}.
\begin{table*}
\begin{center}
\caption{Direct CP violation measurements in \prt{\Dz \to \pi\pi} and
  \prt{\Dz \to \pi pi}, and the average by the Heavy Flavour Averaging
  Group, status January~2009~\cite{hfag}.\label{tab:CPV}}
\begin{tabular}{||c | c *{2}{|r @{$\pm$} c @{$\pm$} l} ||}
\hline\hline
Year & Experiment 
  & \multicolumn{3}{c|}{$A_{CP}$ \prt{\Dz \to \pi \pi}} 
  & \multicolumn{3}{c||}{$A_{CP}$ \prt{\Dz \to KK}} \\
\hline
2008 &	BELLE~\cite{CPVBelle}  & +0.0043 & 0.0052 & 0.0012 
                               & -0.0043 & 0.0030 & 0.0011 \\
2008 &	BaBar~\cite{CPVBabar}  & -0.0024 & 0.0052 & 0.0022
                               & +0.0000 & 0.0034 & 0.0013 \\
2005 &	CDF~\cite{CPVCDF}      & +0.010  & 0.013 & 0.006
                               & +0.020  & 0.012 & 0.006 \\  
2002 &	CLEO~\cite{CPVCLEO}    & +0.019  & 0.032 & 0.008
                               & +0.000  & 0.022 & 0.008 \\ 
2000 &	FOCUS~\cite{CPVFocus}  &  +0.048 & 0.039 & 0.025
                               & -0.001  & 0.022 & 0.015 \\ 
1998 &	E791~\cite{CPVE791}    & -0.049  & 0.078 & 0.030
                               & -0.001  & 0.022 & 0.015 \\
1995 &  CLEO~\cite{Bartelt:1995vr} & \multicolumn{3}{c|}{}
                               & +0.080 & \multicolumn{2}{l||}{0.061} \\
1994 &  E687~\cite{Frabetti:1994kv}  & \multicolumn{3}{c|}{} 
                               & +0.024 & \multicolumn{2}{l||}{ 0.084}\\
                               \hline
\multicolumn{2}{|| l|}{HFAG average~\cite{hfag}} 
                               & +0.0022 & \multicolumn{2}{l|}{0.0037} 
                               & -0.0016 & \multicolumn{2}{l||}{0.0023} 
\\\hline\hline
\end{tabular}
\end{center}
\end{table*}
Results for these two modes, and averages, are listed in
\tabref{tab:CPV}. Other direct CP violation measurements have been
published by BaBar~\cite{Aubert:2008yd, Aubert:2005gj},
BELLE~\cite{Arinstein:2008zh, Tian:2005ik}, CLEO~\cite{cleoDPP2009,
  CLEO:CPVKKpi:2008, absCLEODs, :2007zt, CroninHennessy:2005sy,
  Asner:2003uz, Brandenburg:2001ze, Kopp:2000gv, Bonvicini:2000qm},
FOCUS~\cite{Link:2005th, Link:2001zj}, E791~\cite{Aitala:1996sh}, and
E687~\cite{Frabetti:1994kv}. This includes the recent \cleo\ results
listed in \tabref{tab:DPP_results}. A comprehensive list of results
and averages can found on the HFAG website~\cite{hfag}.  No evidence
for CP violation in the charm sector has emerged yet. It is
interesting to note that the CDF result~\cite{CPVCDF} in
\tabref{tab:CPV} was obtained with only approximately $2\%$ of CDF's
current dataset. A simple scaling of the statistical error suggests that,
if CDF repeated this analysis with the full dataset, the statistical
precision of this single measurement could match the current
world-average. The challenge will of course be to control systematic
uncertainties at a similar level, and there are other reasons why this
simple scaling is too naive, such as the reduction in trigger
efficiency for charm events at higher luminosities. But even with
these caveats, this illustrates the importance and promise of charm
physics at hadron colliders. Most of CDF's charm data have yet to be
analysed, and even larger samples will soon be available at
LHCb~\cite{LHCbCharm}, with the prospect of a rich charm physics
programme with high sensitivity to New Physics.

\section{Conclusions}

Since the last CHARM conference in 2007, large new data samples have
become available and have been analysed, resulting in dramatic
improvements of the precsion of non-leptonic decay rates of charm
mesons, and the discovery of many new decay channels. These are
important parameters in their own right, provide tests of symmetries
of the strong interaction such as U-spin and SU(3), and are set to
provide important input for the analysis of \prt{B} decays, as most B
mesons decay to charm. One of the most sensitive probes for New
Physics is CP violation in the charm sector, which is predicted to be
$<10^{-3}$ in the SM. While at CHARM~2007, the most precise
measurements of direct CP violation achieved a precision at the
percent level, today this has reached the permil level. So far,
however, there has been no evidence for CP violation.

Dedicated charm experiments have unique capabilities, especially when
running at the charm threshold, but are by no means the only source of
charm physics. Many recent measurements have exploited the vast charm
samples at the B factories and CDF. This is an encouraging trend in
view of the start of data taking at LHCb. LHCb will collect
unprecidented charm samples. CDF has shown that precision charm
physics is possible at a hadron collider, and has, for most analyses,
only used a fraction of its dataset. On the other hand, there will
also be new results from the charm threshold with its unique
properties: BES~III is about to take data at the \prt{\psi(3770)}, and
\cleo's dataset continues to be analysed. So the prospects for charm
physics are bright, with continued analyses of $e^+ e^-$ data, new
results from the charm threshold, and enourmous datasets collected at
hadron colliders.

\label{sec:BFresults}
\begin{table*}
  \begin{center}
    \caption{\label{tab:DPP_results} Branching fractions in \prt{D^0
        \to PP} modes measured using \cleo's full dataset, reproduced
      from~\cite{cleoDPP2009}. The table shows the branching ratio
      relative to the normalization modes $D^{0} \rightarrow K^{-}
      \pi^{+}$, $D^{+} \rightarrow K^{-} \pi^{+} \pi^{+}$, and
      $D_{s}^{+} \rightarrow K^{0}_{S} K^{+}$; the resulting Branching
      Fractions; and charge asymmetries
      $\mathcal{A}_{CP}$. Uncertainties are statistical error,
      systematic error, and the error from the input branching
      fractions of normalization modes. (For $D^0$, the normalization
      mode is the sum of $D^0 \rightarrow K^- \pi^+$ and $D^0
      \rightarrow K^+ \pi^-$, see~\cite{cleoDPP2009} for details.)}
    \begin{tabular}{ l c c c}  \hline \hline
      Mode
      & $\BF_{\mathrm{mode}}/\BF_{\mathrm{Normalization}}$ (\%)
      & This result $\BF$ (\%)
      & $\mathcal{A}_{CP}$ (\%) \\ \hline

      $D^{0} \rightarrow K^{+} K^{-}$ & 10.41 $\pm$ 0.11 $\pm$ 0.11 & 0.407 $\pm$ 0.004 $\pm$ 0.004 $\pm$ 0.008 &  \\
      $D^{0} \rightarrow K^{0}_{S} K^{0}_{S}$ & 0.41 $\pm$ 0.04 $\pm$ 0.02 & 0.0160 $\pm$ 0.0017 $\pm$ 0.0008 $\pm$ 0.0003 &  \\
      $D^{0} \rightarrow \pi^{+} \pi^{-}$ & 3.70 $\pm$ 0.06 $\pm$ 0.09 & 0.145 $\pm$ 0.002 $\pm$ 0.004 $\pm$ 0.003 &  \\
      $D^{0} \rightarrow \pi^{0} \pi^{0}$ & 2.06 $\pm$ 0.07 $\pm$ 0.10 & 0.081 $\pm$ 0.003 $\pm$ 0.004 $\pm$ 0.002 &  \\
      $D^{0} \rightarrow K^{-} \pi^{+}$ & 100 & 3.9058 external input~\cite{CLEO:sys} & 0.5 $\pm$ 0.4 $\pm$ 0.9 \\
      $D^{0} \rightarrow K^{0}_{S} \pi^{0}$ & 30.4 $\pm$ 0.3 $\pm$ 0.9 & 1.19 $\pm$ 0.01 $\pm$ 0.04 $\pm$ 0.02 &  \\
      $D^{0} \rightarrow K^{0}_{S} \eta$ & 12.3 $\pm$ 0.3 $\pm$ 0.7 & 0.481 $\pm$ 0.011 $\pm$ 0.026 $\pm$ 0.010 &  \\
      $D^{0} \rightarrow \pi^{0} \eta$ & 1.74 $\pm$ 0.15 $\pm$ 0.11 & 0.068 $\pm$ 0.006 $\pm$ 0.004 $\pm$ 0.001 &  \\
      $D^{0} \rightarrow K^{0}_{S} \eta'$ & 24.3 $\pm$ 0.8 $\pm$ 1.1 & 0.95 $\pm$ 0.03 $\pm$ 0.04 $\pm$ 0.02 &  \\
      $D^{0} \rightarrow \pi^{0} \eta'$ & 2.3 $\pm$ 0.3 $\pm$ 0.2 & 0.091 $\pm$ 0.011 $\pm$ 0.006 $\pm$ 0.002 &  \\
      $D^{0} \rightarrow \eta \eta$ & 4.3 $\pm$ 0.3 $\pm$ 0.4 & 0.167 $\pm$ 0.011 $\pm$ 0.014 $\pm$ 0.003 &  \\
      $D^{0} \rightarrow \eta \eta'$ & 2.7 $\pm$ 0.6 $\pm$ 0.3 & 0.105 $\pm$ 0.024 $\pm$ 0.010 $\pm$ 0.002 &  \\ \hline

      $D^{+} \rightarrow K^{-} \pi^{+} \pi^{+}$ & 100 & 9.1400 external input~\cite{CLEO:sys} & -0.1 $\pm$ 0.4 $\pm$ 0.9 \\
      $D^{+} \rightarrow K^{0}_{S} K^{+}$ & 3.35 $\pm$ 0.06 $\pm$ 0.07 & 0.306 $\pm$ 0.005 $\pm$ 0.007 $\pm$ 0.007 & -0.2 $\pm$ 1.5 $\pm$ 0.9 \\
      $D^{+} \rightarrow \pi^{+} \pi^{0}$ & 1.29 $\pm$ 0.04 $\pm$ 0.05 & 0.118 $\pm$ 0.003 $\pm$ 0.005 $\pm$ 0.003 & 2.9 $\pm$ 2.9 $\pm$ 0.3 \\
      $D^{+} \rightarrow K^{0}_{S} \pi^{+}$ & 16.82 $\pm$ 0.12 $\pm$ 0.37 & 1.537 $\pm$ 0.011 $\pm$ 0.034 $\pm$ 0.033 & -1.3 $\pm$ 0.7 $\pm$ 0.3 \\
      $D^{+} \rightarrow K^{+} \pi^{0}$ & 0.19 $\pm$ 0.02 $\pm$ 0.01 & 0.0172 $\pm$ 0.0018 $\pm$ 0.0006 $\pm$ 0.0004 & -3.5 $\pm$ 10.7 $\pm$ 0.9 \\
      $D^{+} \rightarrow K^{+} \eta$ & $<0.14$ (90\% C.L.) & $<0.013$ (90\% C.L.) &  \\
      $D^{+} \rightarrow \pi^{+} \eta$ & 3.87 $\pm$ 0.09 $\pm$ 0.19 & 0.354 $\pm$ 0.008 $\pm$ 0.018 $\pm$ 0.008 & -2.0 $\pm$ 2.3 $\pm$ 0.3 \\
      $D^{+} \rightarrow K^{+} \eta'$ & $<0.20$ (90\% C.L.) & $<0.018$ (90\% C.L.) &  \\
      $D^{+} \rightarrow \pi^{+} \eta'$ & 5.12 $\pm$ 0.17 $\pm$ 0.25 & 0.468 $\pm$ 0.016 $\pm$ 0.023 $\pm$ 0.010 & -4.0 $\pm$ 3.4 $\pm$ 0.3 \\ \hline

      $D_{s}^{+} \rightarrow K^{0}_{S} K^{+}$ & 100 & 1.4900 external input~\cite{Peterpaper} & 4.7 $\pm$ 1.8 $\pm$ 0.9 \\
      $D_{s}^{+} \rightarrow \pi^{+} \pi^{0}$ & $<2.3$ (90\% C.L.) & $<0.037$ (90\% C.L.) &  \\
      $D_{s}^{+} \rightarrow K^{0}_{S} \pi^{+}$ & 8.5 $\pm$ 0.7 $\pm$ 0.2 & 0.126 $\pm$ 0.011 $\pm$ 0.003 $\pm$ 0.007 & 16.3 $\pm$ 7.3 $\pm$ 0.3 \\
      $D_{s}^{+} \rightarrow K^{+} \pi^{0}$ & 4.2 $\pm$ 1.4 $\pm$ 0.2 & 0.062 $\pm$ 0.022 $\pm$ 0.004 $\pm$ 0.004 & -26.6 $\pm$ 23.8 $\pm$ 0.9 \\
      $D_{s}^{+} \rightarrow K^{+} \eta$ & 11.8 $\pm$ 2.2 $\pm$ 0.6 & 0.176 $\pm$ 0.033 $\pm$ 0.009 $\pm$ 0.010 & 9.3 $\pm$ 15.2 $\pm$ 0.9 \\
      $D_{s}^{+} \rightarrow \pi^{+} \eta$ & 123.6 $\pm$ 4.3 $\pm$ 6.2 & 1.84 $\pm$ 0.06 $\pm$ 0.09 $\pm$ 0.11 & -4.6 $\pm$ 2.9 $\pm$ 0.3 \\
      $D_{s}^{+} \rightarrow K^{+} \eta'$ & 11.8 $\pm$ 3.6 $\pm$ 0.6 & 0.18 $\pm$ 0.05 $\pm$ 0.01 $\pm$ 0.01 & 6.0 $\pm$ 18.9 $\pm$ 0.9 \\
      $D_{s}^{+} \rightarrow \pi^{+} \eta'$ & 265.4 $\pm$ 8.8 $\pm$ 13.9 & 3.95 $\pm$ 0.13 $\pm$ 0.21 $\pm$ 0.23 & -6.1 $\pm$ 3.0 $\pm$ 0.3 \\

     \hline \hline
    \end{tabular}
  \end{center}
\end{table*}

\begin{table*}[tb]
  \begin{center}
    \caption{\label{tab:incDs} \cleo's $D_s$ inclusive yield results
      from ~\cite{CLEO:incDs:2009}. Uncertainties are statistical and
      systematic, respectively. The inclusive $K^{0}_{L}$ results are
      only used as a check for $K^{0}_{S}$. The $D^+_s \rightarrow
      K^{0}_{L} X$ yield requires a correction before comparing with
      the $D^+_s \rightarrow K^{0}_{S} X$ yield, as explained
      in~\cite{CLEO:incDs:2009}. PDG~\cite{pdg2008} averages are
      shown in the last column, when available.  Reproduced
      from~\cite{CLEO:incDs:2009}.}
    \begin{tabular}{ l rclcl l rcl rcl }
      \hline \hline
      Mode
      & \multicolumn{5}{c}{Yield(\%)}
      & ~~~~$K^{0}_{L}$ Mode
      & \multicolumn{3}{c}{Yield(\%)}
      & \multicolumn{3}{r}{$\mathcal{B}$(PDG)(\%)}  \\ \hline

      $D^+_s \rightarrow $$\pi^+ X$ & 119.3 & $\pm$ & 1.2 & $\pm$ & 0.7 & & & & & & &   \\
      $D^+_s \rightarrow $$\pi^- X$ & 43.2 & $\pm$ & 0.9 & $\pm$ & 0.3 & & & & & & &   \\
      $D^+_s \rightarrow $$\pi^0 X$ & 123.4 & $\pm$ & 3.8 & $\pm$ & 5.3 & & & & & & &   \\
      $D^+_s \rightarrow $$K^+ X$ & 28.9 & $\pm$ & 0.6 & $\pm$ & 0.3 & & & & & 20 & $^{+}_{-}$ & $^{18}_{14}$   \\
      $D^+_s \rightarrow $$K^- X$ & 18.7 & $\pm$ & 0.5 & $\pm$ & 0.2 & & & & & 13 & $^{+}_{-}$ & $^{14}_{12}$   \\
      $D^+_s \rightarrow $$\eta X$ & 29.9 & $\pm$ & 2.2 & $\pm$ & 1.7 & & & & & & &   \\
      $D^+_s \rightarrow $$\eta' X$ & 11.7 & $\pm$ & 1.7 & $\pm$ & 0.7 & & & & & & &   \\
      $D^+_s \rightarrow $$\phi X$ & 15.7 & $\pm$ & 0.8 & $\pm$ & 0.6 & & & & & & &   \\
      $D^+_s \rightarrow $$\omega X$ & 6.1 & $\pm$ & 1.4 & $\pm$ & 0.3 & & & & & & &   \\
      $D^+_s \rightarrow $$f_0(980) X, f_0(980) \rightarrow \pi^+\pi^-$ &  \multicolumn{5}{c}{ $<$ 1.3\% (90\%  CL) } & & & & & & &   \\
      $D^+_s \rightarrow $$K^{0}_{S} X$ &  19.0 & $\pm$ & 1.0 & $\pm$ & 0.4  & $D^+_s \rightarrow $$K^{0}_{L} X$  &  15.6 & $\pm$ & 2.0 &  20 & $\pm$ & 14   \\
      $D^+_s \rightarrow $$K^{0}_{S} K^{0}_{S} X$ & 1.7 & $\pm$ & 0.3 & $\pm$ & 0.1 & $D^+_s \rightarrow $$K^{0}_{L} K^{0}_{S} X$ &  5.0 & $\pm$ & 1.0 & & &   \\
      $D^+_s \rightarrow $$K^{0}_{S} K^+ X$ & 5.8 & $\pm$ & 0.5 & $\pm$ & 0.1 & $D^+_s \rightarrow $$K^{0}_{L} K^+ X$ &  5.2 & $\pm$ & 0.7 & & & \\
      $D^+_s \rightarrow $$K^{0}_{S} K^- X$ & 1.9 & $\pm$ & 0.4 & $\pm$ & 0.1 & $D^+_s \rightarrow $$K^{0}_{L} K^- X$ &  1.9 & $\pm$ & 0.3 & & & \\
      $D^+_s \rightarrow $$K^+ K^- X$ & 15.8 & $\pm$ & 0.6 & $\pm$ & 0.3 & & & & & & &   \\
      $D^+_s \rightarrow $$K^+ K^+ X$ &  \multicolumn{5}{c}{ $<$ 0.26\% (90\%  CL) } & & & & & & &   \\
      $D^+_s \rightarrow $$K^- K^- X$ &  \multicolumn{5}{c}{ $<$ 0.06\% (90\%  CL) } & & & & & & &   \\
      \hline \hline
    \end{tabular}
  \end{center}
\end{table*}      

\bibliographystyle{h-physrev}
\bibliography{bibliography}


\end{document}